\documentclass[10pt]{article}
\usepackage{geometry} 
\usepackage{graphicx}
\usepackage{subfig}
\usepackage{amssymb}
\usepackage{amsmath}
\usepackage{hyperref}
\usepackage{epstopdf}
\usepackage{caption}
\newcommand{\vecw}{\boldsymbol{\mathbf{w}}}
\newcommand{\vecx}{\boldsymbol{\mathbf{x}}}

\newcommand{\bv}{\boldsymbol{\beta}}

\usepackage[T1]{fontenc}
\usepackage[utf8]{inputenc}
\usepackage{authblk}

\title{Selective recruitment designs for improving observational studies using electronic health records}
\author[1]{James E. Barrett\thanks{Contact: james.barrett@kcl.ac.uk}\thanks{These authors contributed equally to this work.}}
\author[2]{Aylin Cakiroglu$^\dag$}
\author[3]{Catey Bunce}
\author[4,5,6]{Anoop Shah}
\author[4,5]{Spiros Denaxas}
\affil[1]{Cancer Cell Biology \& Imaging, King's College London, London, SE1 1UL, U.K.}
\affil[2]{The Francis Crick Institute, London, NW1 1AT, U.K.}
\affil[3]{Division of Health \& Social Care Research, King's College London, London, SE1 1UL, U.K.}
\affil[4]{UCL Institute of Health Informatics, University College London, Gower Street, London WC1E 6BT, U.K.}
\affil[5]{Health Data Research U.K., London, U.K.}
\affil[6]{University College London Hospitals NHS Trust, 250 Euston Road, London NW1 2PG, U.K.}
\date{\today}

\begin{document}
\maketitle

\begin{abstract}
Large scale electronic health records (EHRs) present an opportunity to quickly identify suitable individuals in order to directly invite them to participate in an observational study. EHRs can contain data from millions of individuals, raising the question of how to optimally select a cohort of size $n$ from a larger pool of size $N$. In this paper we propose a simple selective recruitment protocol that selects a cohort in which covariates of interest tend to have a uniform distribution. We show that selectively recruited cohorts potentially offer greater statistical power and more accurate parameter estimates than randomly selected cohorts. Our protocol can be applied to studies with multiple categorical and continuous covariates. We apply our protocol to a numerically simulated prospective observational study using an EHR database of stable acute coronary disease patients from 82,089 individuals in the U.K. Selective recruitment designs require a smaller sample size, leading to more efficient and cost-effective studies.
\end{abstract}

%=================================================%
%
%
\section{Introduction}
\label{sec:intro}
%
%
%=================================================%

Large scale electronic health records present the possibility of conducting prospective observational studies by directly identifying individuals that meet pre-specified criteria \cite{Effoe2017,Cowie2017}. EHRs typically contain clinical covariates and phenotypes that can be linked to laboratory tests, primary and secondary care records, as well as molecular data. In a conventional observational study investigators typically wait for potential recruits to arrive at designated study centres --- a process that can take years to complete, if at all \cite{Carlisle2015}. EHRs may potentially contain millions of patients and in many cases there will be an abundance of eligible patients for a particular study. EHRs offer the obvious advantages of faster recruitment and reduced costs but they also raise the interesting question of how to optimally select a cohort of $n$ individuals from a pool of size $N$ where $N \gg n$. 

The aim of an observational study is to establish a statistical relationship between covariates and clinical outcomes of interest. We assume that the covariates of interest are available in the EHR database, but that the outcomes are not, either because they are not routinely recorded or because more detailed or rigorous measurements are required. EHRs present an opportunity to select patients on the basis of their covariates in order to invite them to participate in the study. The simplest selection strategy is to randomly select $n$ individuals from the pool. As we shall see this generally wont't provide the greatest statistical power. An alternative strategy is to preferentially select a more ``informative'' cohort, where informativeness is defined in terms of covariate values. In this paper we propose a simple strategy that attempts to form a cohort in which each covariate has a uniform distribution. Each member of the pool is assigned a recruitment probability. Individuals that will contribute to a uniform cohort distribution are deemed more informative, and consequently will have a higher probability of recruitment.

To gain some intuition for this idea consider several patients with identical covariate values compared to several patients with slightly different covariate values. Although both groups are informative, the latter patients are inherently more informative because they tell us how the outcome depends on different values of the covariates. Our selective recruitment strategy means we are less likely to make repeated observations of similar individuals, and more likely to explore the covariate space efficiently. Statistical inference is based on observed regularities between covariates and outcomes. It is therefore advantageous to acquire observations evenly throughout the covariate space rather than a concentration of data points within a restricted region of the space.

As a further example, consider a pool population with a single binary covariate coded as $+1$ and $-1$. Selecting a cohort with an equal number of $+1$ and $-1$ observations will maximise statistical power. From a statistical perspective there is no \emph{a priori} justification for selecting more of one covariate value than the other, even if the covariate is unequally distributed in the population. The desire for an a priori uniform covariate distribution in our cohort reflects Keynes' \emph{principle of indifference}\cite{keynes1922treatise} which states that ``equal probabilities must be assigned to each of several arguments if there is an absence of positive ground for assigning unequal ones''.

The ability to be selective about which patients to invite onto a study is only possible with the emergence of large-scale EHRs. While the clinical utility of EHRs is increasingly recognised \cite{coorevits2013electronic,jensen2012mining,murdoch2013inevitable,weiskopf2013methods} the underlying infrastructure is still developing and the use of EHRs for research purposes is fraught with issues such as missing and incomplete data, data quality, accuracy, confidentiality, interoperability, security and patient consent. These problems have been discussed in depth in the literature \cite{jensen2012mining,coorevits2013electronic,hripcsak2013next,weiskopf2013methods}, and we will restrict our focus to statistical issues relating to the use of EHRs as a recruitment aid. An example of EHR based recruitment is the European Electronic Health Record systems for Clinical Research (EHR4CR) platform \cite{murphy2010serving}.

The remainder of this paper is organised as follows. In Section 2 we review previous work on controlling the distribution of covariates in a clinical study. We describe our selective recruitment protocol in Section 3. In Section 4 we perform numerical simulations and study the operating characteristics of our protocol in comparison to randomised selection strategies. In Section 5, as a proof of concept, we apply our protocol to a numerically simulated observational study based on EHR data from 82,089 patients with stable acute coronary disease in the U.K. We discuss our findings in Section 6 and present our conclusions in Section 7.

%=================================================%
%
%
\section{Background}
\label{sec:background}
%
%
%=================================================%

The concept of controlling the covariate distribution within a study cohort has previously been implemented in a  variety contexts. These techniques share a common theme: creating a favourable distribution of covariates in order to increase statistical power and reduce the risk of bias. The most straightforward approach is \emph{stratified sampling} in which the population is divided into distinct strata, out of which individuals are randomly sampled \cite{levy2013sampling}. This ensures distinct subpopulations are equally represented. \emph{Matching} is a technique that can be applied retrospectively to observational datasets containing an \emph{exposure} (or treatment) group and a \emph{control} group \cite{rubin1973matching}. A subset of the data are selected as a control group such that the distribution of covariates within the exposure and control group are as similar as possible. Both groups are therefore more comparable and estimates of group differences are less prone to bias.

When the exposure and control groups don't match perfectly a parametric model can be used to account for differences in covariates \cite{cochran1973controlling}. When there are a large number of covariates it becomes difficult to form a matching cohort and instead \emph{propensity score matching} can be used \cite{rosenbaum1983central}. Matching methods can be viewed as a means to reduce model dependent bias \cite{ho2007matching}. This is because the parametric model used to adjust for covariate imbalances may be misspecified in practice and with matched groups the dependence on model assumptions is diminished.  All matching methods are prone to bias when unmeasured covariates are associated with the outcome of interest and it is frequently assumed that all relevant covariates are measured (although this is impossible to verify in reality).

Covariate balancing methods have also been used in the theory of experimental design. \emph{Stratified blocking} designs randomise treatment and controls within predefined strata \cite{fisher1935design} thus ensuring both groups are similar in terms of the stratified covariates. Covariate-adaptive clinical trials allocate patients onto treatment arms in a manner that tries to minimise the covariate imbalance between arms \cite{taves1974minimization,pocock1975sequential,lin2015pursuit}. Another field that uses covariate information to select samples is \emph{active machine learning}. The aim is to actively seek data points that are anticipated to be informative. There are various ways to define informativeness \cite{settles2010active}. For example, individuals that are expected to reduce the posterior entropy or reduce future prediction errors are deemed more informative. Several of these concepts were previously applied to selective recruitment trial designs\cite{barrett16}.

%=================================================%
%
%
\section{Methods}
\label{sec:methods}
%
%
%=================================================%

We assume that each individual in the pool is characterised by a $d$-dimensional vector of covariates $\vecx$, and denote the clinical outcome of interest as $y$. We will consider both binary and time-to-event outcomes in this paper. It is further assumed that $y$ is unavailable in the EHR system, either because it is not routinely measured or requires further measurements. In this paper we will focus on selecting a cohort for a prospective observational study in which the goal is to establish the statistical relationship between $\vecx$ and $y$.

Choosing a uniform distribution to reflect the absence of prior knowledge is similar in spirit to the use of \emph{uninformative priors} in Bayesian inference \cite{gelman2014bayesian}. One potential problem with uninformative priors is that they depend on how a covariate is defined. A uniform distribution over height, for instance, will not correspond to a uniform distribution over body mass index (which is based on the square of height). Some uninformative priors have been developed that are invariant to re-parameterisation of a covariate such as Jeffery's prior \cite{jeffreys1946invariant}. For the purposes of this article we will assume that covariates have been appropriately defined in advance and use uniform distributions to reflect a lack of prior knowledge. 

\subsection{Selective recruitment with a single binary covariate}

Our goal is to select a subset of $n$ individuals from within a larger pool of $N$ individuals. The vector $\vecx$ consists of either categorical or continuous covariates. We denote binary clinical outcomes by $y\in\{-1,+1\}$. Our first strategy is to select individuals such that the distribution of covariates across the cohort is as close to uniform as possible. Suppose we have a single binary covariate $x\in\{-1,+1\}$ and that the proportion of individuals in the pool with $x=+1$ is $p$. We can recruit individual $i$ from the pool with probability
\begin{equation}
\rho(x_i) = \left\{\begin{array}{cc}
(1-p)/c & \text{if $x_i=+1$} \\
p/c & \text{if $x_i=-1$} \\
\end{array}
\right.
\quad \text{for $i=1,\ldots,N$}
\label{eq:binary}
\end{equation}
where the normalisation constant is $c=\sum_{i=1}^N \rho(x_i)$. This recruitment strategy will ensure that on average the covariate is uniformly distributed within the cohort.

\begin{figure}[t]
\centering
\begin{tabular}{c c c}
\subfloat[Pool distribution]{\includegraphics[scale=0.185]{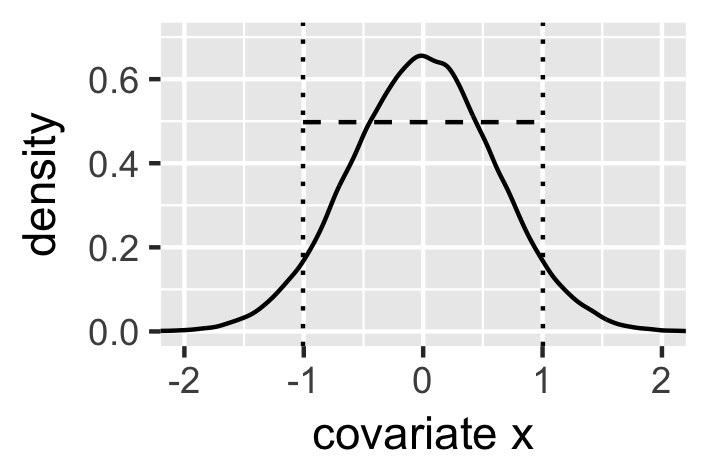}} & \subfloat[Recruitment probability]{\includegraphics[scale = 0.185]{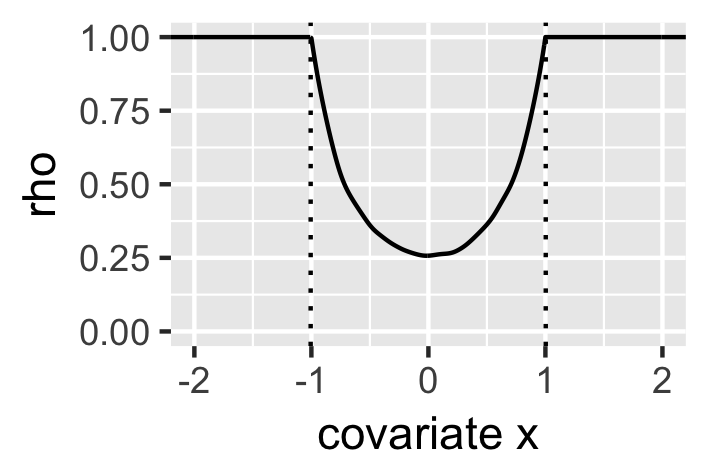}}& \subfloat[Cohort distribution]{\includegraphics[scale = 0.185]{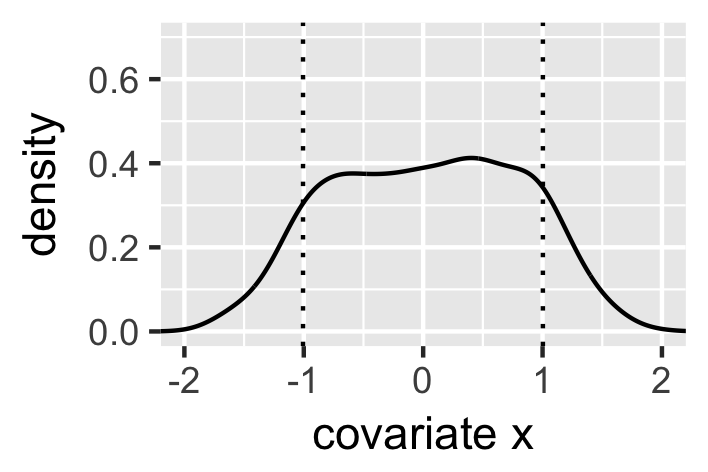}}\\
\end{tabular}\caption{In (a) is the pool distribution ($N=100,000$) of a single covariate $x$ (solid black line). The two vertical dotted lines correspond to the 0.05 and 0.95 quantiles. The horizontal dashed line corresponds to the value of $q$ (as defined in (\ref{eq:continuous})). In (b) is the recruitment probability as a function of $x$. In (c) is the cohort distribution ($n=1,000$) after selective recruitment from the pool.}
\label{fig:continuous}
\end{figure}

\subsection{Selective recruitment with a single continuous covariate}
\label{sec:continuous}

A covariate $x\in\mathbb{R}$ with infinite support means that selecting a cohort with a uniform distribution is not possible. As a pragmatic compromise we attempt to form a uniform cohort distribution between the 0.05 and 0.95 quantiles of the pool distribution (denoted by $x_l$ and $x_u$ respectively). We first generate an empirical density estimate $p(x)$ of the pool distribution. A recruitment probability for an individual with covariate $x_i$ is given by
\begin{equation}
\rho(x_i) = \left\{\begin{array}{cc}
\frac{1}{c}\frac{q}{c' p(x_i)} & \text{if $x_l\leq x_i \leq x_u$} \\
\frac{1}{c} & \text{otherwise} \\
\end{array}
\right.
\quad \text{for $i=1,\ldots,N$}
\label{eq:continuous}
\end{equation}
where $q=1/(x_u-x_l)$. The constants $c$, defined as above, and $c' = \text{max}_{x_l\leq x \leq x_u} q/p(x)$ ensure the probabilities are appropriately normalised. An example of this can be seen in Figure \ref{fig:continuous} (b).

\subsection{Selective recruitment with multiple covariates}

\begin{figure}[t]
\centering
\begin{tabular}{c c c}
\subfloat[Pool distribution]{\includegraphics[scale=0.475]{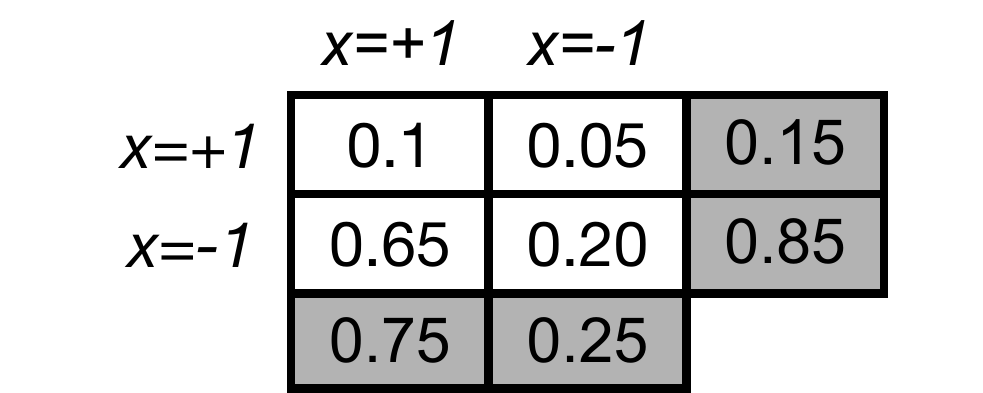}} & \subfloat[Marginally balanced]{\includegraphics[scale = 0.475]{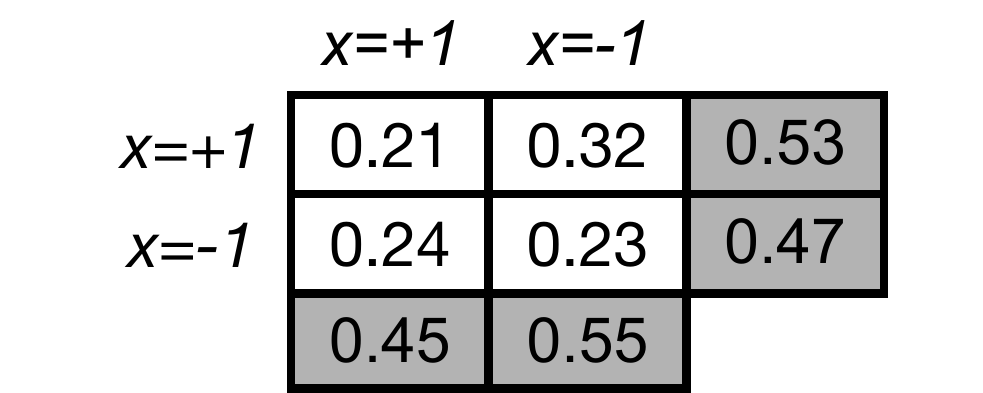}}& \subfloat[Jointly balanced]{\includegraphics[scale = 0.475]{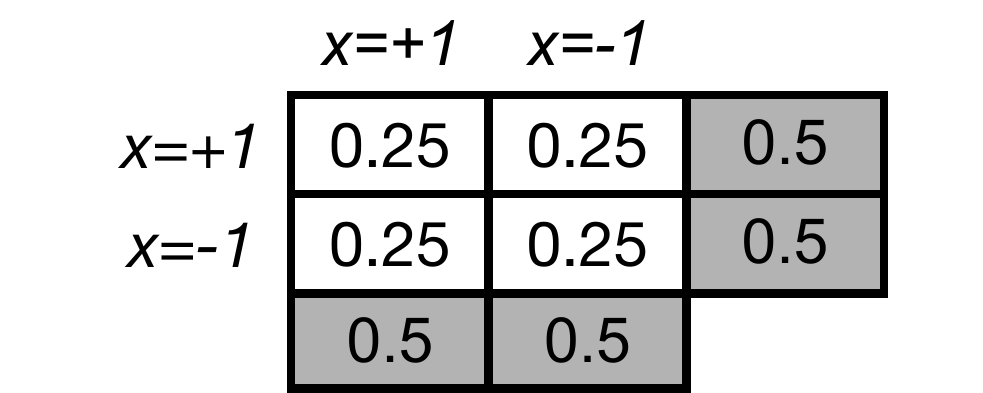}}\\
\end{tabular}
\caption{In (a) is the pool distribution of two binary covariates. In (b) is the cohort distribution after applying (\ref{eq:multivariate}) (and assuming large $N$ and $n$). In (c) is a cohort with a perfectly balanced joint distribution.}
\label{fig:multivariate}
\end{figure}

When we have $d$ covariates one option is to try and balance the marginal distribution of each covariate. This can be achieved by
\begin{equation}
\rho(x_i) = \frac{1}{c}\prod_{\mu=1}^d \rho_{\mu}(x_i)
\label{eq:multivariate}
\end{equation}
where $\rho_{\mu}(x_i)$ is given by either (\ref{eq:binary}) or (\ref{eq:continuous}). An example of this protocol with two binary covariates is shown in Figure \ref{fig:multivariate} (b). An alternative strategy when all covariates are binary is to simply balance the joint distribution of covariates within the cohort (as in Figure \ref{fig:multivariate} (c)). This is the preferred method and can always be achieved by simply stratifying the pool into four groups and randomly selecting the requisite number of individuals from each group. However, when the pool size is relatively small and the number of covariates in a study is large this this generally won't be possible and the marginally balanced method may be used instead.

%=================================================%
%
%
\section{Results from numerical simulation studies}
\label{sec:numerical}
%
%
%=================================================%

In order to assess the performance of these different selection protocols we performed several numerical simulations. We evaluated the statistical power, mean square error, and type I error rates under various conditions.

\subsection{Binary covariates}

A pool of $N=10,000$ individuals with two binary covariates was generated from the distribution shown in Figure \ref{fig:multivariate} (a). We recruited $n$ individuals from the pool according to three different protocols, marginally balanced (Figure \ref{fig:multivariate} (b)), jointly balanced (Figure \ref{fig:multivariate} (c)), and random selection. Binary outcomes $y=\pm1$ were generated according to a logistic regression model $p(y=+1|\vecx) = 1/(1+\text{exp}(-w_0-\vecw\cdot\vecx))$ with parameters set to $w_0=-1/6$ and $\vecw=(1/3,+1/3)$. For each cohort of size $n$ a logistic regression model was fitted and statistical power was calculated as the proportion of inferred parameters that were statistically significantly at $\alpha=0.05$. Statistical power and the mean square error between true and inferred parameter values as a function of cohort size $n$ is plotted in Figure \ref{fig:binary-balance}. Selective recruitment offers a clear advantage with little difference between the jointly and marginally balanced protocols. We also found that the Type I error rates in cohorts formed using the different protocols were all well controlled at the expected 5\% error rate (Supplementary Figure 1). The existence of unmeasured covariate introduces a bias to the parameter estimates but this bias is independent of the cohort distribution (Supplmentary Figure 2).

\begin{figure}[t]
\centering
\begin{tabular}{c c c}
\subfloat[Statistical power]{\includegraphics[scale=0.7]{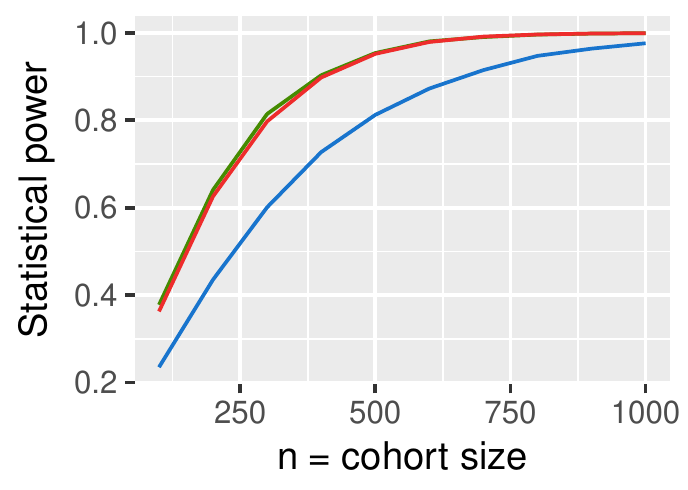}} & \subfloat[Mean square error]{\includegraphics[scale = 0.7]{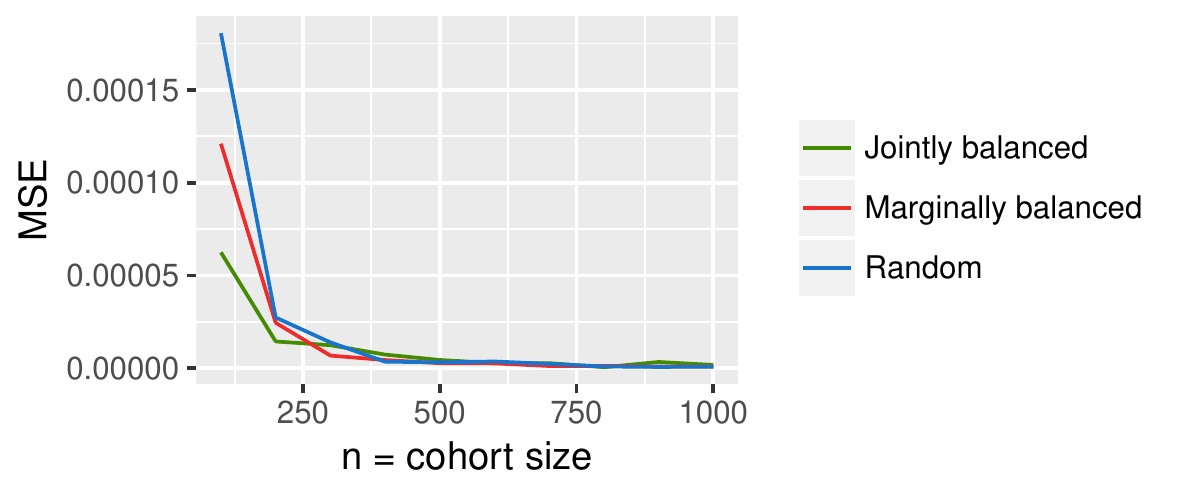}}\\
\end{tabular}
\caption{Statistical power and mean square error as a function of cohort size in the case of two binary covariates.}
\label{fig:binary-balance}
\end{figure}

\subsection{Continuous covariate}

\begin{figure}[t]
\centering
\begin{tabular}{c c c}
\subfloat[Statistical power]{\includegraphics[scale=0.7]{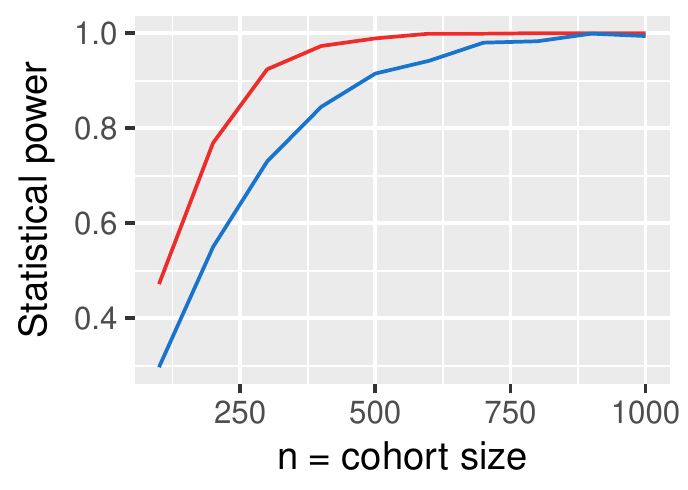}} & \subfloat[Mean square error]{\includegraphics[scale = 0.7]{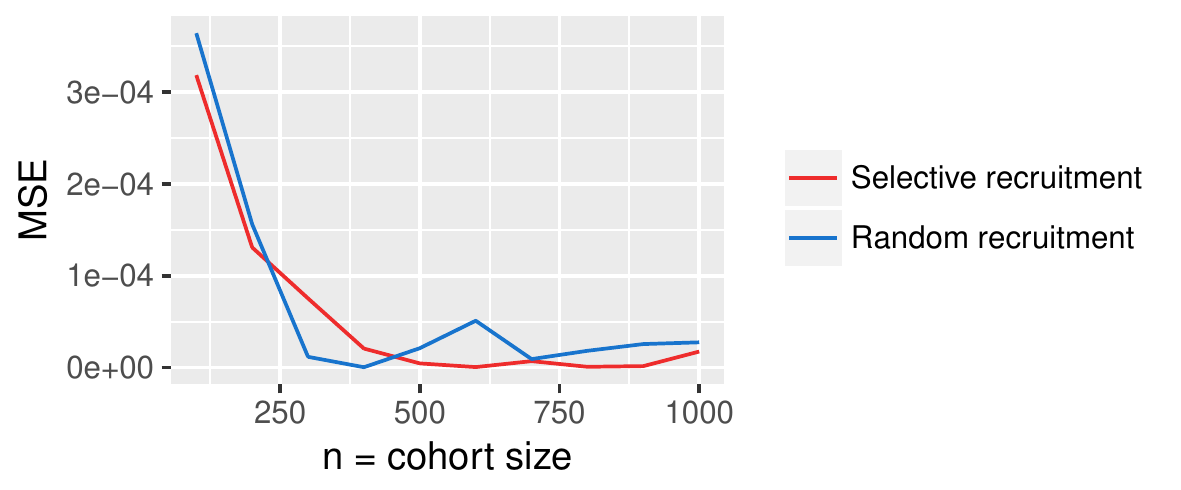}}\\
\end{tabular}
\caption{Statistical power and mean square error as a function of cohort size for the case of one continuous covariate.}
\label{fig:continuous-balance}
\end{figure}

A pool of $N=10,000$ individuals was generated with a single normally distributed covariate $x$ with zero mean and standard deviation 0.608 (such that the 0.05 and 0.95 quantiles are equal to $-1$ and $+1$ for convenience). Cohorts were selected according to (\ref{eq:continuous}) and compared to a randomised recruitment design. A logistic regression model with parameters $w_0=-1/2$ and $w=-1/4$ was used to generate outcomes. The statistical power and mean square error between true and inferred parameters, obtained after fitting logistic regression models to each simulated cohort, are plotted in Figure \ref{fig:continuous-balance}. We find that the selective recruitment protocol offers a clear gain in in statistical power. For example, to achieve a power of 90\% approximately 275 individuals would need to be recruited using a selective recruitment design in comparison to approximately 500 individuals in a randomised design.

%=================================================%
%
%
\section{Results from application to a cardiovascular EHR database}
\label{sec:caliber}
%
%
%=================================================%

In order to demonstrate how a selective-recruitment protocol can be used in practice we simulated a prospective observational study using an EHR database of 82,089 anonymised patients with stable coronary artery disease from the CALIBER resource\cite{CALIBER2012,CALIBER2014} (described below). The data consist of 30 biomarkers and risk factors and the primary outcome was time-to-death (all-cause mortality). Our aim was to select a cohort of $n=1,000$ individuals and study the associations between the 30 covariates and time-to-death. We compared the operating characteristics of randomly and selectively recruited cohorts.

For the purposes of our proof-of-concept simulation both covariates and the outcome of interest are already available. In practice, however, a prospective observational study would be required in situations where the desired outcome was unavailable or situations where a study with more rigorous and detailed measurements were required. In these situations EHR resources could potentially be used for the recruitment of individuals onto a study in which the clinical outcome of interest would subsequently be measured. The type of study we are simulating is similar to the Cardiovascular Health Study which was a prospective observational study aiming to establish cardiovascular risk factors associated with five year mortality in a population of 5,201 adults in the United States\cite{fried1998risk}. We propose that instead of slowly accruing 5,201 individuals at designated study centres, a cohort instead could be formed using EHRs, should they be available. The results above below that a smaller (but more informative) cohort could potentially offer the same level of power as a randomly recruited cohort.

\subsection{Data sources}
\label{sec:data}

CALIBER was established to provide access to longitudinal data of linked electronic health records through the creation of a common data model with reproducible phenotypes and metadata. Patients were linked across three clinical data sources: the Clinical Practice Research Datalink (CPRD), Hospital Episodes Statistics (HES), and cause-specific mortality (from the Office of National Statistics). CPRD provides information about anthropometric measurements, laboratory tests, clinical diagnoses, prescriptions, and medical procedures, coded with the Read controlled clinical terminology. The primary care practices in CPRD and the subset of linked practices used in the present analysis are representative of the UK primary care setting and have been validated for epidemiological research \cite{herrett2013,herrett2015}. HES provides information about diagnoses (coded with the tenth revision of the International Classification of Diseases statistical classification system) and interventional procedures related to all elective and emergency hospital admissions across all National Health Service hospitals in England. 

The eligible patients were chosen from a cohort of a previous study on stable coronary artery disease prediction using CALIBER data \cite{rapsomaniki2013}. All variables that were chosen as predictors in the previous study were used as covariates in our simulation. These included age, diabetes, smoking, systolic blood pressure, diastolic blood pressure, total cholesterol, HDL cholesterol, serum creatinine, haemoglobin, total white blood cell count, CABG or PCI surgery within six months prior to study entry, abdominal aortic aneurysm prior to study entry, index of multiple deprivation, ethnicity, hypertension diagnosis or medication prior to study entry, use of long acting nitrates prior to study entry, diabetes diagnosis prior to study entry, peripheral arterial disease prior to study entry, and history of depression, anxiety disorder, cancer, renal disease, chronic obstructive pulmonary disease, atrial fibrillation, or stroke. We excluded the history of MI and liver disease because both were highly correlated with other covariates in our dataset. A summary of the patient population used in this study is shown in Supplementary Table 1. Dichotomous covariates were coded as $-1$ or $+1$. Continuous covariates were linearly scaled such that the 0.05 and 0.95 quantiles are equal to $-1$ and $+1$ respectively. Full details of covariates, study population definitions, and an overview and details of the imputation methods can be found in the Section 2 of the Supplementary material.

\subsection{Simulation of a prospective observational study using the CALIBER dataset }

The pool of available patients was split into ten smaller pools each containing 8,208 individuals. Splitting the pool into ten smaller pools allows us to run ten independent simulations and average the results. From each pool a cohort of 1,000 patients was selected either at random or according to the selective recruitment protocol.  At the end of each simulation we fitted a Cox proportional hazards model and recorded which covariates were found to be statistically significant at $\alpha=0.05$. These results were compared to a Cox model fitted to the full dataset of 82,089 patients. We found in our simulations that in the full dataset 27 out of 30 covariates were found to be statistically significant. Of these 27 we found that on average 9 were statistically significant using the selective recruitment protocol compared to an average of 6.8 when using a random protocol. An average of 0.4 and 0.2 of the 3 covariates which were not found to be significant in the full dataset were found to be significant in the selectively and randomly recruited cohorts respectively. The mean square difference between inferred model parameters in the selectively recruited cohorts and full dataset was 0.02 compared to 0.21 for randomly selected cohorts. 

\begin{figure}
\centering
\includegraphics[scale=0.6]{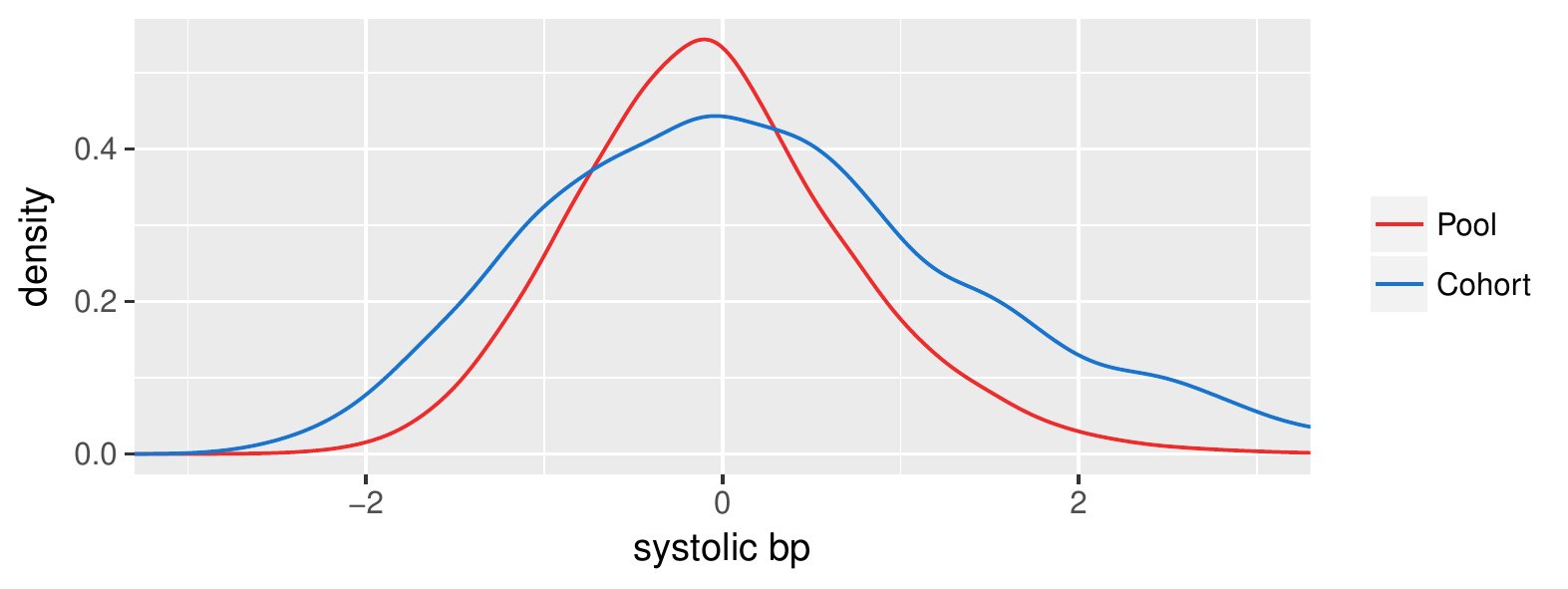}
\caption{The empirical density of systolic blood pressure in a selectively recruited cohort of size 1,000 compared to the pool of size 82,089.}
\label{fig:systolic}
\end{figure}

An obvious limitation here is that the parameters based on the full dataset are only estimators and not the true parameter values (which are unknown). Nevertheless, given the large size of the dataset ($N=82,089$) relative to the number of covariates ($d=30$) the estimated parameters will be reasonably accurate for the purposes of comparison to estimates based on a small subset ($n=1,000$) of patients. The distribution of covariates within the selectively recruited cohorts was more balanced than the randomly selected cohorts. For each dichotomous covariate we computed the ratio of the less frequent covariate value to the more frequent value. The median value of this ratio in the selectively recruited cohorts was 0.32 compared to 0.13 in the randomly selected cohorts. In Figure \ref{fig:systolic} the empirical cohort density of systolic blood pressure is plotted for one instance of a selectively recruited cohort and compared to the pool density. The covariate has a broader distribution than the pool. Further figures are available in Supplementary Figure 3.

%=================================================%
%
%
\section{Discussion}
\label{sec:disc}
%
%
%=================================================%

When a pool of potential recruits for a study is available, it may be beneficial to preferentially select a study cohort with a maximally informative distribution of covariates. We have shown that preferential selection can lead to greater statistical power for a given sample size. This is achieved by aiming for a cohort in which covariates have a distribution that is as close to uniform as possible. We have shown that our selective recruitment protocol outperforms random selection in terms of power, sample size, and mean square error between true and inferred parameters in numerical simulations. Furthermore, we demonstrated the feasibility of our strategy by simulating realistic prospective observational studies using the CALIBER resource, an EHR with 82,089 patients. A similar study has previously been conducted in the U.S. and our results indicate that using EHR resources to selectively recruit patients would result in smaller sample size requirements.

EHRs offer a potentially useful recruitment aid for clinical studies. A medical centre could use a local database of patients in order to identify patients with a particular condition for the purposes of a study. National level EHRs could help to identify patients with rare conditions, and help to form a cohort with a favourable composition. The techniques considered here may also be applicable to the recruitment of patients for clinical trials. It was previously shown that in trials with biomarkers it may be advantageous to select cohorts that have statistically desirable biomarker distributions \cite{barrett16,barrett17}. We have restricted our present analysis to observational studies but an extension to randomised trials will be considered in future work. Another application of the protocol proposed here is to the cohort selection of a follow-up study to a clinical trial. In such scenarios a subset of patients are typically followed over a longer time period in order to acquire further evidence and monitor for adverse side effects. Here too, selective recruitment methods may be useful for selecting the maximally informative subset of individuals for the follow-up study.

In previous selective recruitment designs more sophisticated measures of informativeness were used to select study participants\cite{barrett16}. Various techniques were considered to estimate informativeness such as the expected decrease in posterior entropy and the expected decrease in prediction error or variance. The protocol proposed here could be extended to increase the recruitment probability of individuals that are deemed to be highly informative. Such approaches are sensitive to the choice of statistical model however. For instance, previous research found that in a logistic regression model or a proportional hazards model individuals with extreme covariate values are deemed most informative since effect sizes are implicitly assumed to be most pronounced in these individuals. Further research will be required to assess the characteristics of such selection protocols.

Besides the obvious logistical advantages of EHR based recruitment, preferential selection of informative cohorts will reduce the overall sample size requirements leading to more cost-effective studies. We anticipate that in the future the prospect of leveraging EHRs to boost recruitment will become increasingly attractive.

%=================================================%
%
%
\section{Conclusion}
%
%
%=================================================%

Electronic health records present an opportunity to select a subset of individuals from a larger pool for the purposes of a clinical study. Rather than randomly selecting a cohort, preferentially composing a cohort with an informative covariate distribution may offer increased statistical power, lower mean square error, and smaller sample size requirements without compromising the type I error rate.

%=================================================%
%
%
\section{Acknowledgements}
%
%
%=================================================%

This work was supported by Health Data Research UK, which receives its funding from HDR UK Ltd (NIWA1) funded by the UK Medical Research Council, Engineering and Physical Sciences Research Council, Economic and Social Research Council, Department of Health and Social Care (England), Chief Scientist Office of the Scottish Government Health and Social Care Directorates, Health and Social Care Research and Development Division (Welsh Government), Public Health Agency (Northern Ireland), British Heart Foundation (BHF) and the Wellcome Trust.
This study was supported by National Institute for Health Research (RP-PG-0407-10314), Wellcome Trust (086091/Z/08/Z). This study was supported by the Farr Institute of Health Informatics Research at UCL Partners, from the Medical Research Council, Arthritis Research UK, British Heart Foundation, Cancer Research UK, Chief Scientist Office, Economic and Social Research Council, Engineering and Physical Sciences Research Council, National Institute for Health Research, National Institute for Social Care and Health Research, and Wellcome Trust (MR/K006584/1).

This study was approved by the  Medicines and Healthcare Products Regulatory Agency (MHRA) Independent Scientific Advisory Committee (ISAC) - protocol reference: 17\_032.

This study is based in part on data from the Clinical Practice Research Datalink obtained under licence from the UK Medicines and Healthcare products Regulatory Agency. The data is provided by patients and collected by the NHS as part of their care and support. The interpretation and conclusions contained in this study are those of the authors alone.

Hospital Episode Statistics Copyright (2019), re-used with the permission of The Health \& Social Care Information Centre. All rights reserved. The OPCS Classification of Interventions and Procedures, codes, terms and text is Crown copyright (2016) published by Health and Social Care Information Centre, also known as NHS Digital and licensed under the Open Government Licence available at \url{http://www.nationalarchives.gov.uk/doc/open-government-licence/version/3/}

This study was carried out as part of the CALIBER programme. CALIBER, led from the UCL Institute of Health Informatics, is a research resource consisting of linked electronic health records phenotypes, methods and tools, specialised infrastructure, and training and support (\url{//www.ucl.ac.uk/health-informatics/caliber}).

\bibliographystyle{unsrt}
\bibliography{refs}

\begin{thebibliography}{10}

\bibitem{Effoe2017}
Valery~S Effoe, Jeffrey~A Katula, Julienne~K Kirk, Carolyn~F Pedley, Linda~Y
  Bollhalter, W~Mark Brown, Margaret~R Savoca, Stedman~T Jones, Janet Baek, and
  Alain~G Bertoni.
\newblock The use of electronic medical records for recruitment in clinical
  trials: findings from the lifestyle intervention for treatment of diabetes
  trial.
\newblock {\em Trials}, 17(1):496, 2016.

\bibitem{Cowie2017}
Martin~R Cowie, Juuso~I Blomster, Lesley~H Curtis, Sylvie Duclaux, Ian Ford,
  Fleur Fritz, Samantha Goldman, Salim Janmohamed, J{\"o}rg Kreuzer, Mark
  Leenay, et~al.
\newblock Electronic health records to facilitate clinical research.
\newblock {\em Clinical Research in Cardiology}, 106(1):1--9, 2017.

\bibitem{Carlisle2015}
Benjamin Carlisle, Jonathan Kimmelman, Tim Ramsay, and Nathalie MacKinnon.
\newblock Unsuccessful trial accrual and human subjects protections: an
  empirical analysis of recently closed trials.
\newblock {\em Clinical Trials}, 12(1):77--83, 2015.

\bibitem{keynes1922treatise}
John~Maynard Keynes.
\newblock A treatise on probability, 1922.

\bibitem{coorevits2013electronic}
Pascal Coorevits, M~Sundgren, Gunnar~O Klein, A~Bahr, B~Claerhout, C~Daniel,
  M~Dugas, D~Dupont, A~Schmidt, P~Singleton, et~al.
\newblock Electronic health records: new opportunities for clinical research.
\newblock {\em Journal of internal medicine}, 274(6):547--560, 2013.

\bibitem{jensen2012mining}
Peter~B Jensen, Lars~J Jensen, and S{\o}ren Brunak.
\newblock Mining electronic health records: towards better research
  applications and clinical care.
\newblock {\em Nature Reviews Genetics}, 13(6):395--405, 2012.

\bibitem{murdoch2013inevitable}
Travis~B Murdoch and Allan~S Detsky.
\newblock The inevitable application of big data to health care.
\newblock {\em Jama}, 309(13):1351--1352, 2013.

\bibitem{weiskopf2013methods}
Nicole~Gray Weiskopf and Chunhua Weng.
\newblock Methods and dimensions of electronic health record data quality
  assessment: enabling reuse for clinical research.
\newblock {\em Journal of the American Medical Informatics Association},
  20(1):144--151, 2013.

\bibitem{hripcsak2013next}
George Hripcsak and David~J Albers.
\newblock Next-generation phenotyping of electronic health records.
\newblock {\em Journal of the American Medical Informatics Association},
  20(1):117--121, 2013.

\bibitem{murphy2010serving}
Shawn~N Murphy, Griffin Weber, Michael Mendis, Vivian Gainer, Henry~C Chueh,
  Susanne Churchill, and Isaac Kohane.
\newblock Serving the enterprise and beyond with informatics for integrating
  biology and the bedside (i2b2).
\newblock {\em Journal of the American Medical Informatics Association},
  17(2):124--130, 2010.

\bibitem{levy2013sampling}
Paul~S Levy and Stanley Lemeshow.
\newblock {\em Sampling of populations: methods and applications}.
\newblock John Wiley \& Sons, 2013.

\bibitem{rubin1973matching}
Donald~B Rubin.
\newblock Matching to remove bias in observational studies.
\newblock {\em Biometrics}, pages 159--183, 1973.

\bibitem{cochran1973controlling}
William~G Cochran and Donald~B Rubin.
\newblock Controlling bias in observational studies: A review.
\newblock {\em Sankhy{\=a}: The Indian Journal of Statistics, Series A}, pages
  417--446, 1973.

\bibitem{rosenbaum1983central}
Paul~R Rosenbaum and Donald~B Rubin.
\newblock The central role of the propensity score in observational studies for
  causal effects.
\newblock {\em Biometrika}, pages 41--55, 1983.

\bibitem{ho2007matching}
Daniel~E Ho, Kosuke Imai, Gary King, and Elizabeth~As Stuart.
\newblock Matching as nonparametric preprocessing for reducing model dependence
  in parametric causal inference.
\newblock {\em Political analysis}, 15(3):199--236, 2007.

\bibitem{fisher1935design}
Ronald~A Fisher.
\newblock The design of experiments. 1935.
\newblock {\em Oliver and Boyd, Edinburgh}, 1935.

\bibitem{taves1974minimization}
Donald~R Taves.
\newblock Minimization: a new method of assigning patients to treatment and
  control groups.
\newblock {\em Clinical pharmacology and therapeutics}, 15(5):443, 1974.

\bibitem{pocock1975sequential}
Stuart~J Pocock and Richard Simon.
\newblock Sequential treatment assignment with balancing for prognostic factors
  in the controlled clinical trial.
\newblock {\em Biometrics}, pages 103--115, 1975.

\bibitem{lin2015pursuit}
Yunzhi Lin, Ming Zhu, and Zheng Su.
\newblock The pursuit of balance: An overview of covariate-adaptive
  randomization techniques in clinical trials.
\newblock {\em Contemporary clinical trials}, 45:21--25, 2015.

\bibitem{settles2010active}
Burr Settles.
\newblock Active learning literature survey.
\newblock {\em University of Wisconsin, Madison}, 52(55-66):11, 2010.

\bibitem{barrett16}
James~E Barrett.
\newblock Information-adaptive clinical trials: a selective recruitment design.
\newblock {\em Journal of the Royal Statistical Society: Series C (Applied
  Statistics)}, 65(5):797--808, 2016.

\bibitem{gelman2014bayesian}
Andrew Gelman, John~B Carlin, Hal~S Stern, and Donald~B Rubin.
\newblock {\em Bayesian data analysis}, volume~2.
\newblock Chapman \& Hall/CRC Boca Raton, FL, USA, 2014.

\bibitem{jeffreys1946invariant}
Harold Jeffreys.
\newblock An invariant form for the prior probability in estimation problems.
\newblock In {\em Proceedings of the Royal Society of London a: mathematical,
  physical and engineering sciences}, volume 186, pages 453--461. The Royal
  Society, 1946.

\bibitem{CALIBER2012}
Spiros~C Denaxas, Julie George, Emily Herrett, Anoop~D Shah, Dipak Kalra,
  Aroon~D Hingorani, Mika Kivimaki, Adam~D Timmis, Liam Smeeth, and Harry
  Hemingway.
\newblock Data resource profile: Cardiovascular disease research using linked
  bespoke studies and electronic health records (caliber).
\newblock {\em International Journal of Epidemiology}, 41(6):1625--1638, 2012.

\bibitem{CALIBER2014}
Katherine~I. Morley, Joshua Wallace, Spiros~C. Denaxas, Ross~J. Hunter,
  Riyaz~S. Patel, Pablo Perel, Anoop~D. Shah, Adam~D. Timmis, Richard~J.
  Schilling, and Harry Hemingway.
\newblock Defining disease phenotypes using national linked electronic health
  records: A case study of atrial fibrillation.
\newblock {\em PLOS ONE}, 9(11):1--10, 11 2014.

\bibitem{fried1998risk}
Linda~P Fried, Richard~A Kronmal, Anne~B Newman, Diane~E Bild, Maurice~B
  Mittelmark, Joseph~F Polak, John~A Robbins, Julius~M Gardin, Cardiovascular
  Health Study Collaborative~Research Group, et~al.
\newblock Risk factors for 5-year mortality in older adults: the cardiovascular
  health study.
\newblock {\em Jama}, 279(8):585--592, 1998.

\bibitem{herrett2013}
Emily Herrett, Anoop~Dinesh Shah, Rachael Boggon, Spiros Denaxas, Liam Smeeth,
  Tjeerd van Staa, Adam Timmis, and Harry Hemingway.
\newblock Completeness and diagnostic validity of recording acute myocardial
  infarction events in primary care, hospital care, disease registry, and
  national mortality records: cohort study.
\newblock {\em Bmj}, 346:f2350, 2013.

\bibitem{herrett2015}
Emily Herrett, Arlene~M Gallagher, Krishnan Bhaskaran, Harriet Forbes, Rohini
  Mathur, Tjeerd van Staa, and Liam Smeeth.
\newblock Data resource profile: clinical practice research datalink (cprd).
\newblock {\em International journal of epidemiology}, 44(3):827--836, 2015.

\bibitem{rapsomaniki2013}
Eleni Rapsomaniki, Anoop Shah, Pablo Perel, Spiros Denaxas, Julie George, Owen
  Nicholas, Ruzan Udumyan, Gene~Solomon Feder, Aroon~D. Hingorani, Adam Timmis,
  Liam Smeeth, and Harry Hemingway.
\newblock Prognostic models for stable coronary artery disease based on
  electronic health record cohort of 102 023 patients.
\newblock {\em European Heart Journal}, 35(13):844--852, 2014.

\bibitem{barrett17}
James~E Barrett.
\newblock Information-adaptive clinical trials with selective recruitment and
  binary outcomes.
\newblock {\em Statistics in Medicine}, 36(18):2803--2813, 2017.
\newblock sim.7353.

\bibitem{mice}
Stef {van Buuren} and Karin Groothuis-Oudshoorn.
\newblock {mice}: Multivariate imputation by chained equations in r.
\newblock {\em Journal of Statistical Software}, 45(3):1--67, 2011.

\end{thebibliography}

%\bibliographystyle{plainnat}
%\bibliography{refs}
%=================================================%
\newpage
\section*{Supplementary Information}
%=================================================%

\appendix

%=================================================%
\section{Numerical simulations}
%=================================================%

\subsection{Type I error rate}

A pool of $N=10,000$ individuals with two binary covariates was generated from the distribution shown in Figure 2 (a) of the main text. Binary outcomes $y=\pm1$ were generated according to a logistic regression model $p(y=+1|\vecx) = 1/(1+\text{exp}(-w_0-\vecw\cdot\vecx))$. The parameters were set to $w_0=-1/6$ and $\vecw=(=1/3,0)$. Cohorts of size $n$ were selecting around to the marginally balanced, jointly balanced, and random selection protocols. For each cohort a logistic regression model was fitted. The null hypothesis of no association between covariates and outcomes is true for the second component of $\vecw$. In order to estimate the type I error rate we calculated the proportion of times this parameter was inferred to be significant at a 0.05 significance level. The results are shown in Supplementary Figure \ref{fig:typeI}. The type I error rate is well controlled under all selection protocols.

\begin{figure}[h]
\centering
\includegraphics[scale=0.75]{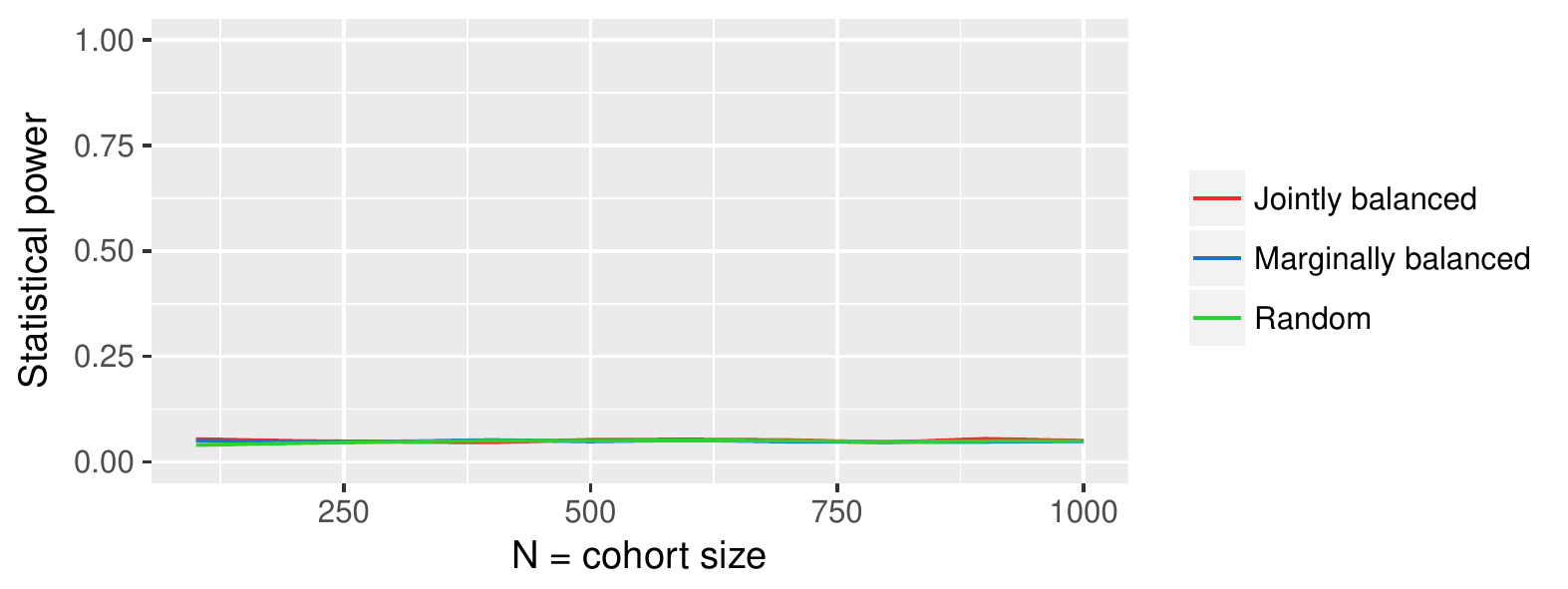}
\caption{Type I error rate as a function of cohort size for the case of two binary covariates.}
\label{fig:typeI}
\end{figure}

\subsection{Unmeasured covariates}

A pool of $N=10,000$ individuals with two binary covariates was generated from the distribution shown in Figure 2 (a) of the main text. Binary outcomes $y=\pm1$ were generated according to a logistic regression model $p(y=+1|\vecx) = 1/(1+\text{exp}(-w_0-\vecw\cdot\vecx))$. The parameters were set to $w_0=-1/6$ and $\vecw=(=-1/3,1/4)$. After the outcomes had been generated the second covariate was removed from the pool (that is, it was an unmeasured covariate). The marginal distribution of the remaining covariate was $p(x_1=1)=0.75$ and $p(x_1=-1)=0.25$.

Cohorts of size $n$ were selected from the pool according to the jointly balanced, and random selection protocols. For each cohort a logistic regression model was fitted. The mean square error between the inferred and true parameter $w_1$ is shown in Supplementary Figure \ref{fig:binary-balance}. The unmeasured covariate introduces a bias into the parameter estimate but this is the same for both types of recruitment protocols.

For comparison a second experiment was run in which a pool of $N=10,000$ individuals with a single binary covariate were generated. A logistic regression model with $w_0=-1/6$ and $w=-1/3$ was used to generate outcomes. Cohorts of size $n$ were selected as above. The mean square error between inferred and true parameter values is also plotted in Supplementary Figure \ref{fig:binary-balance}. Due to the absence of unmeasured covariates there is no bias in this case.

\begin{figure}[h]
\centering
\includegraphics[scale=0.75]{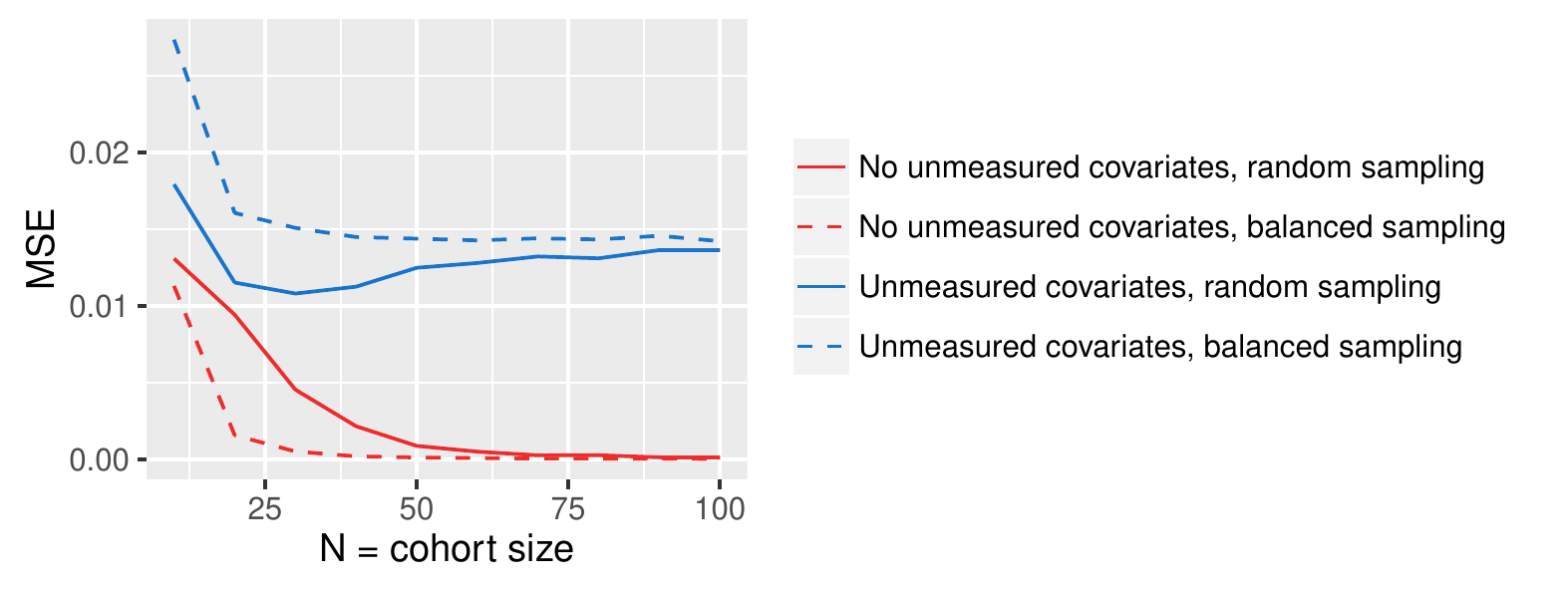}
\caption{Mean square error in the presence and absence of unmeasured covariates.}
\label{fig:binary-balance}
\end{figure}

%=================================================%
\section{The CALIBER dataset}
%=================================================%

\subsection{Study population}

We selected patients from 225 primary care practices registered between January 1997, and March 2010, who fulfilled the following criteria which are aligned with previous work described in \cite{rapsomaniki2013}:
\begin{itemize}
\item Are 18 years or older at study start date or turn 18 during the study period time
\item
\begin{itemize}
\item
Stable angina, defined by Read codes in CPRD for angina diagnosis, positive ischaemia tests, coronary angiogram results recorded or repeat prescriptions for nitrates, or in HES by hospitalisations with a primary spell diagnosis ICD10 code I20.1, I20.8, or I20.9
\item
Myocardial Infarction non-fatal, as defined by Read codes in CPRD or ICD10 I21?I22 as the primary diagnosis in HES
\item
had other coronary artery disease (CHD) in CPRD/HES data, i.e. coronary artery bypass graft (CABG), or percutaneous coronary intervention (PCI)
\item
Unstable angina, defined by Read codes in CPRD or hospital admission with ICD10 code I20.0
\end{itemize}
\end{itemize}

Patients with prior acute events were classified as stable if they survived longer than 6 months after the acute event, and only entered the cohort at this point following definitions in \cite{rapsomaniki2013}. Patients entered on the date of the first myocardial infarction after study eligibility (minimum of one year after registration in a contributing GP practice).

\subsection{Endpoints}

The primary endpoint was all-cause mortality as defined in ONS or CPRD. Patients were censored at the earliest date among death date, relocation to a new primary care practice, or study end date (25 March 2010) following definitions in \cite{rapsomaniki2013}.

\subsection{Imputation}

Multiple imputation was implemented using multivariate imputation by chained equations in the R package mice \cite{mice}. Imputation models were estimated separately for men and women using all 115,305 patients before exclusion criteria were applied (MI or death before study eligibility) and included:

\begin{itemize}
\item
All the baseline covariates used in the main analysis (age, diabetes, smoking, systolic blood pressure, diastolic blood pressure, total cholesterol, HDL cholesterol, body mass index, serum creatinine, haemoglobin, total white blood cell count, CABG or PCI surgery within 6 months prior to study entry, abdominal aortic aneurysm prior to study entry, index of multiple deprivation, ethnicity, hypertension diagnosis or medication prior to study entry, use of long acting nitrates prior to study entry, diabetes diagnosis prior to study entry, peripheral arterial disease prior to study entry, and quadratic age.

\item
Prior (between 1 and 2 years before study entry) and post (between 0 and 2 years after study entry) averages of continuous main analysis covariates and other measurements not in the main analysis (Hba1c, eFGR-CKDEPI, lymphocyte counts, neutrophil counts, eosinophil counts, monocyte counts, basophil counts, platelet counts, pulse pressure).

\item

Coexisting medical conditions (history of heart attack, depression, anxiety disorder, cancer, renal disease, liver disease, chronic obstructive pulmonary disease, atrial fibrillation, or stroke prior to study entry)

\item

The Nelson-Aalen hazard and the event status for each endpoint analyzed in the data.

\end{itemize}

Since many of the continuous variables were non-normally distributed, we log-transformed all continuous variables for imputation and exponentiated back to their original scale for analysis. Only one multiply imputed dataset was generated since any imputation errors are not expected to have a significant effect on our analyses in respect to the comparison of different designs. The distributions of observed and imputed values of all variables followed similar distributions indicating the plausibility of the imputation.

\begin{table}[]
    \centering
    \begin{tabular}{|l|l|}
    \hline
         Characteristic &  Summary \\
         \hline
    Patient population &82,217\\
Age & 68 (47 -- 87) years\\
Gender & 47,016 (57$\%$) men, 35,201 (43$\%$) women\\
Social deprivation (IMD score) &16 (4 -- 52)\\
Diagnosis & 57.4$\%$ stable angina\\
&   11.8$\%$ unstable angina\\
& 13.9$\%$ ST-elevated myocardial infarction\\
& 14.9$\%$ non-ST-elevated myocardial infarction\\
& 1.9$\%$ other coronary heart disease\\
PCI in last 6 months&4.6$\%$\\
CABG in last 6 months&1.9$\%$\\
Abdominal aortic aneurysm & 1.1\%\\
Use of nitrates&26.8$\%$\\
Smoking status &20.0$\%$ current\\
& 35.5$\%$ ex-smoker \\
& 44.3$\%$ non-smoker\\
Hypertension&88.1$\%$\\
Diabetes mellitus&15.4$\%$\\
Total cholesterol&5.18 (3.2 -- 7.76) mmol/l\\
High density lipoprotein cholesterol&1.35 (0.80 -- 2.1) mmol/l\\
Heart failure&9.0$\%$\\
Renal disease & 0.67\%\\
Peripheral arterial disease&7.0$\%$\\
Atrial fibrillation&11.6$\%$\\
Stroke&5.4$\%$\\
Chronic kidney disease&6.7$\%$\\
Chronic obstructive pulmonary disease&35.1$\%$\\
Cancer&8.0$\%$\\
Depression at diagnosis&19.1$\%$\\
Anxiety at diagnosis&12.0$\%$\\
Heart rate&73.03 (51.2 -- 99.3) bpm\\
Diastolic blood pressure & 79.42 (60.0 -- 99.6) mmHg\\
Systolic blood pressure & 140.52 (110.0 -- 178.0) mmHg\\
Creatinine&99.8 (64.8 -- 149.0) mol/l\\
White cell count&7.4 (4.45 -- 11.3) $10^9$/l\\
Haemoglobin&13.74 (11.0 -- 16.64) g/dL\\
Endpoint & 18,930 (23\%) dead, 63,287 (77\%) censored\\
\hline
    \end{tabular}
    \caption{Cohort summary of all variables. Values are quoted to two significant figures and may not sum due to rounding. Continuous values are summarised as median (5th -- 95th percentile).}
    \label{tab:CohortSummary}
\end{table}

\begin{figure}
\centering
\begin{tabular}{c c}
\subfloat[]{\includegraphics[scale=0.4]{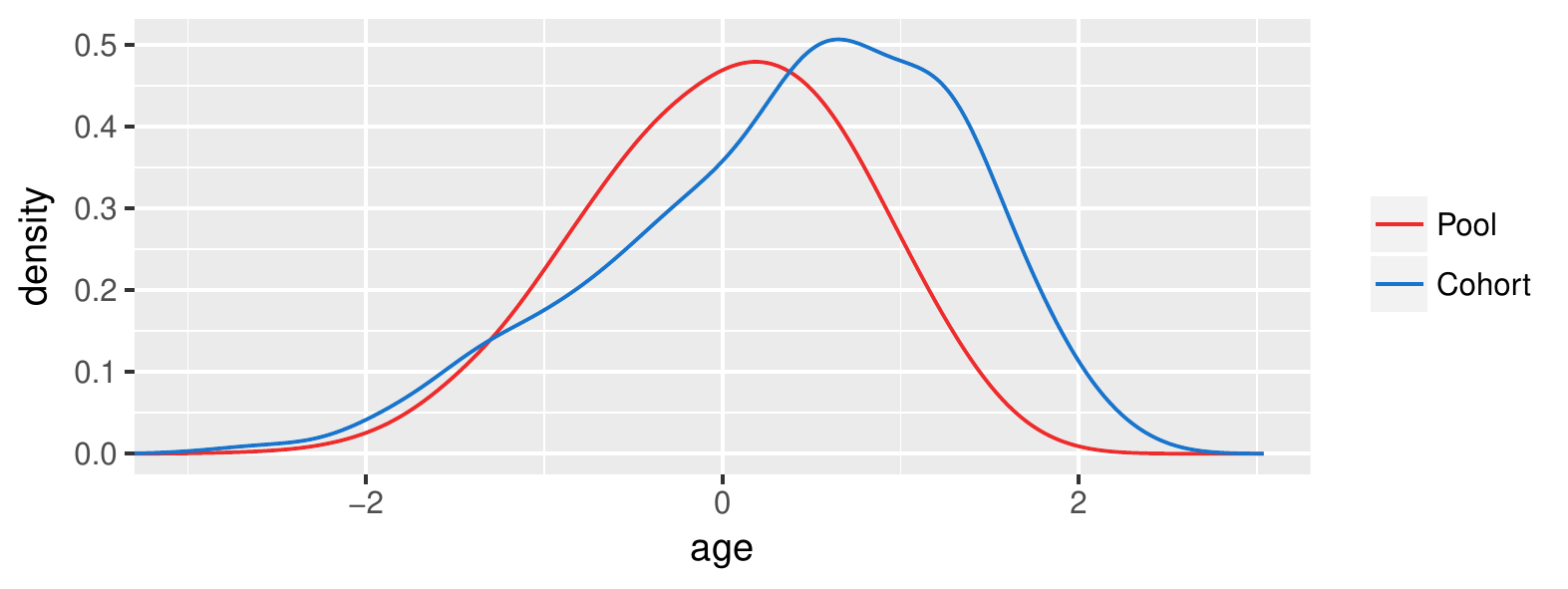}} & \subfloat[]{\includegraphics[scale = 0.4]{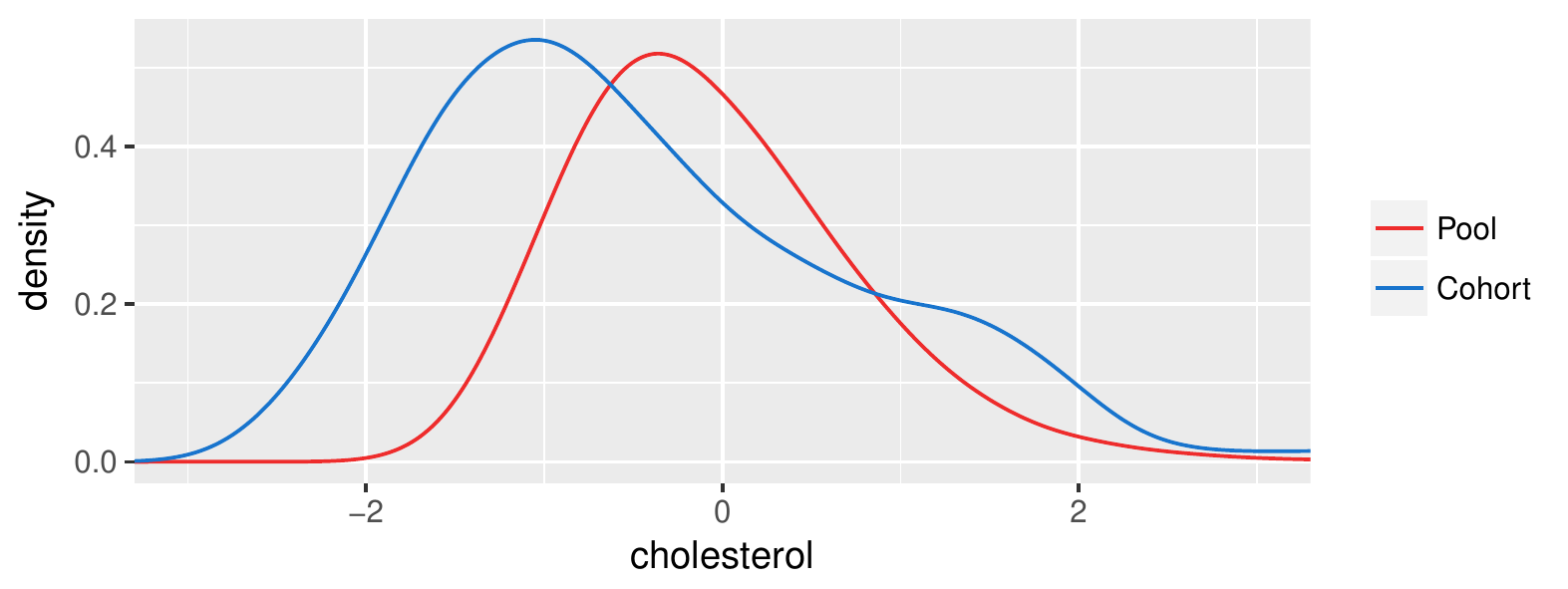}}\\
\subfloat[]{\includegraphics[scale=0.4]{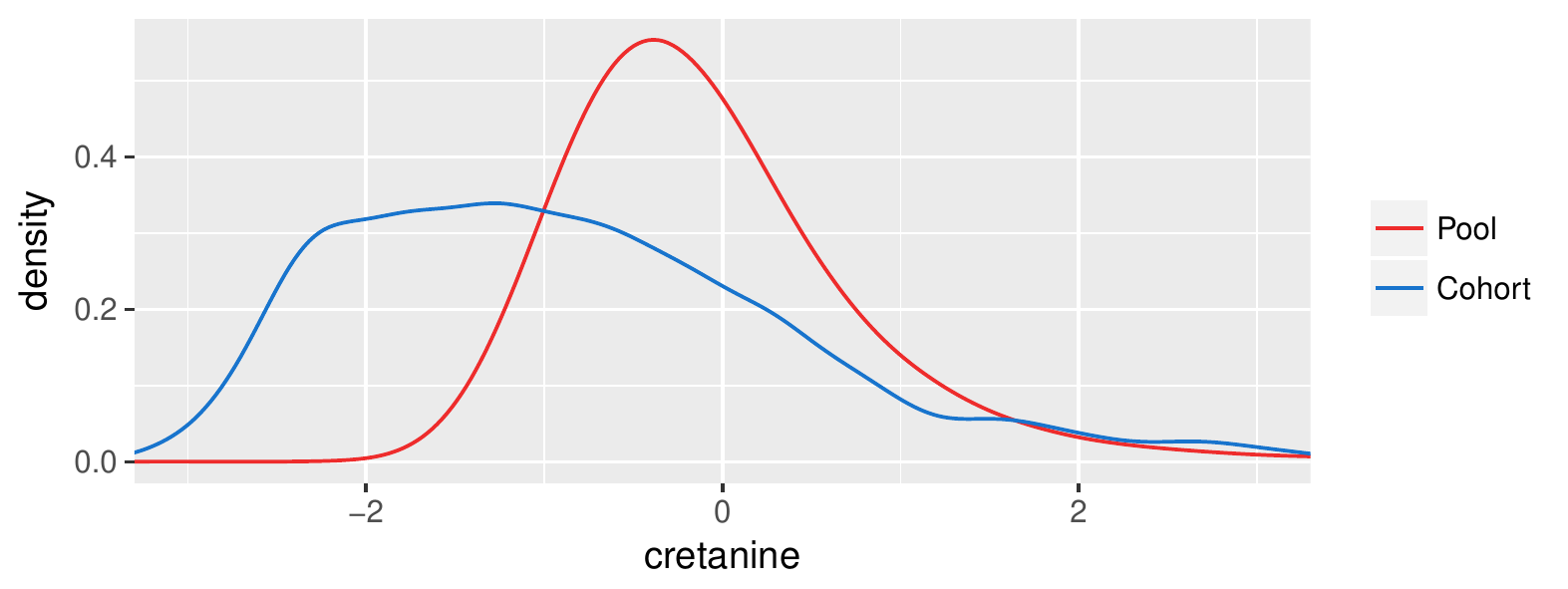}} & \subfloat[]{\includegraphics[scale = 0.4]{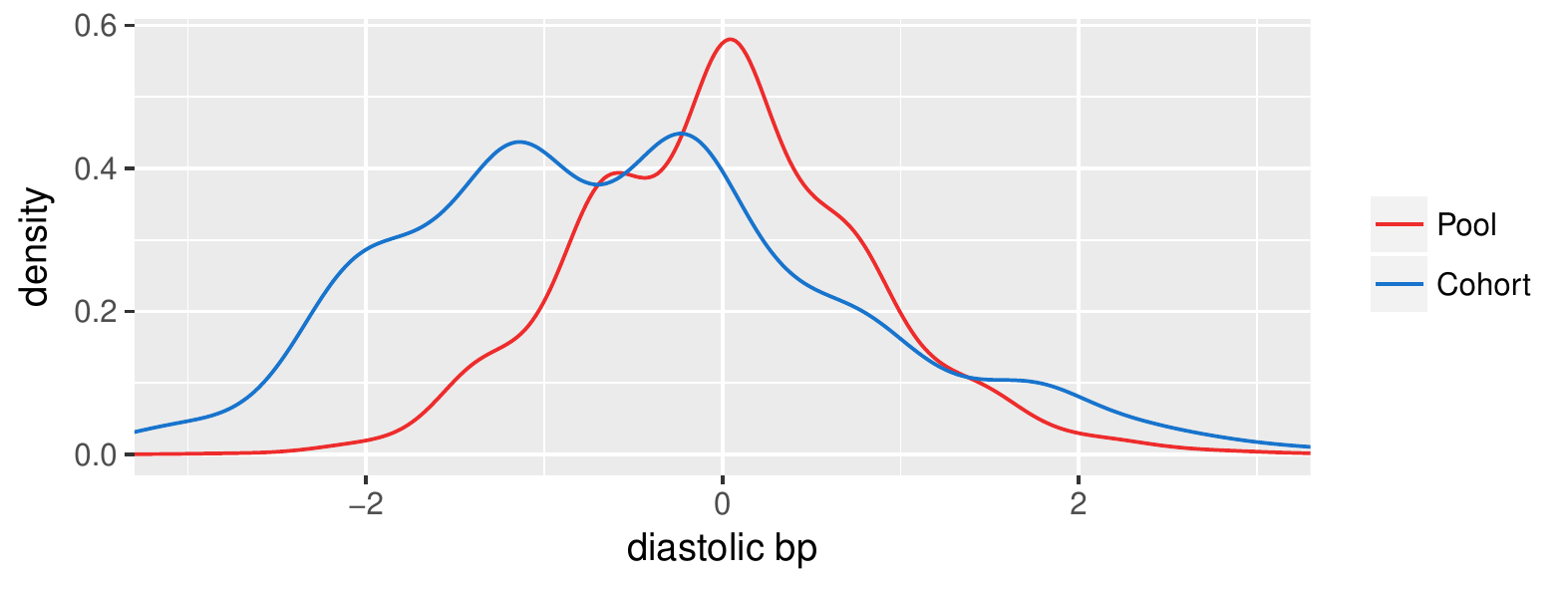}}\\
\subfloat[]{\includegraphics[scale=0.4]{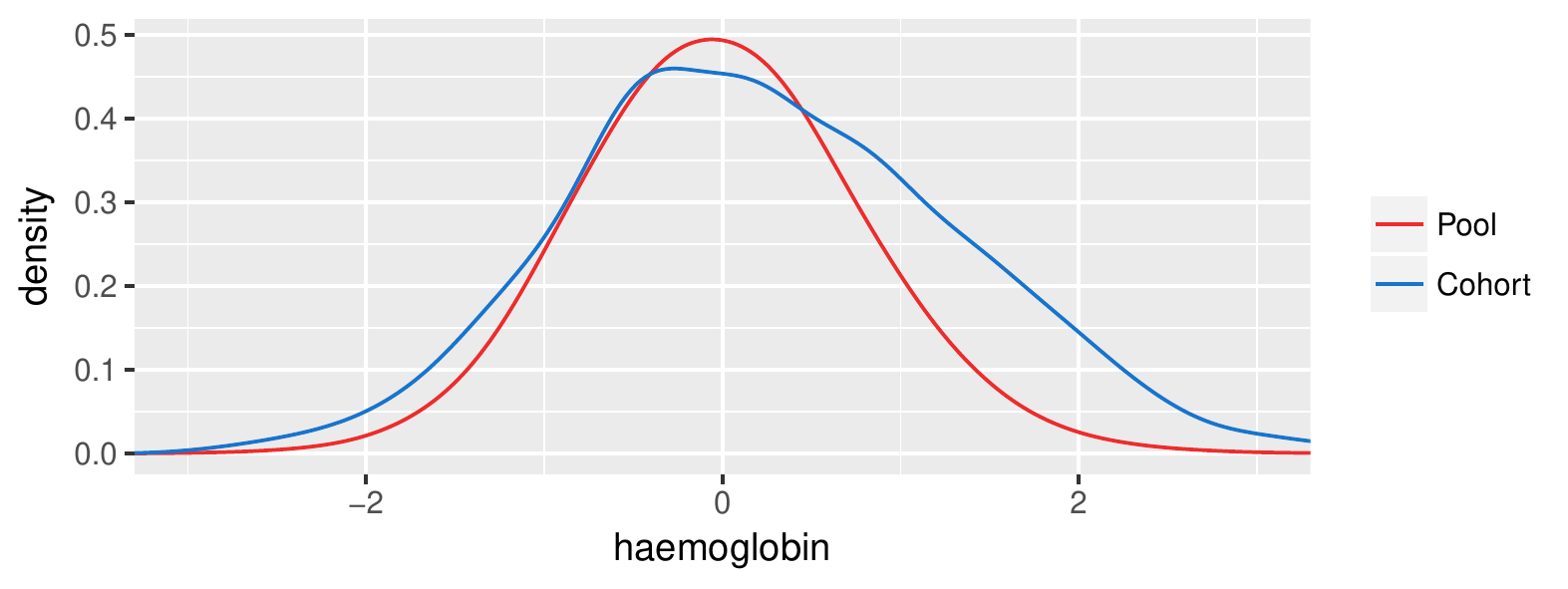}} & \subfloat[]{\includegraphics[scale = 0.4]{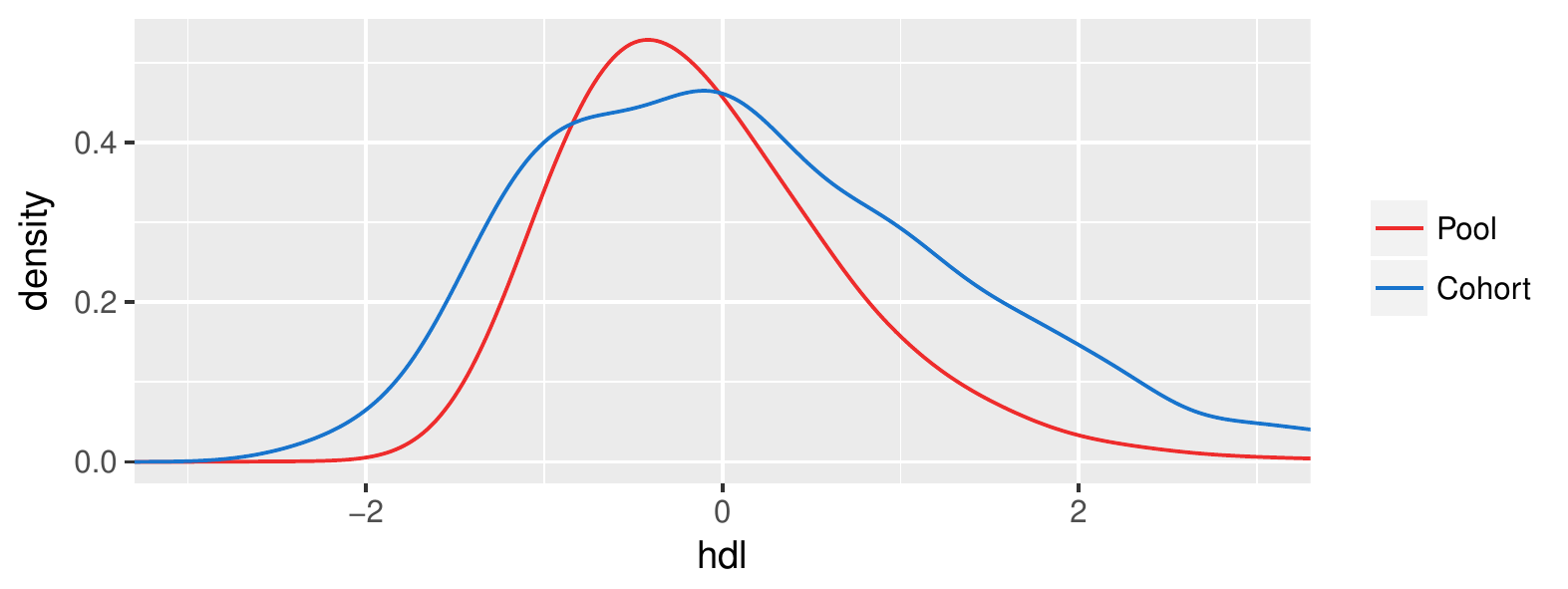}}\\
\subfloat[]{\includegraphics[scale=0.4]{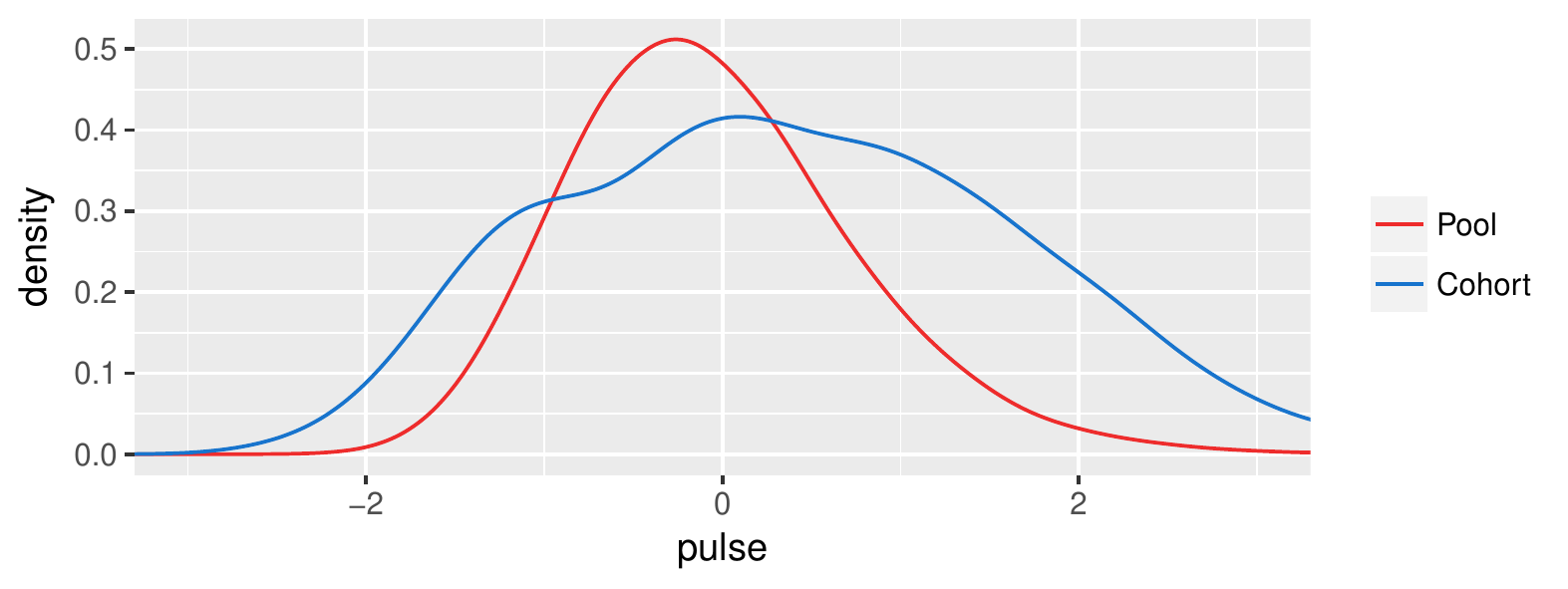}} & \subfloat[]{\includegraphics[scale = 0.4]{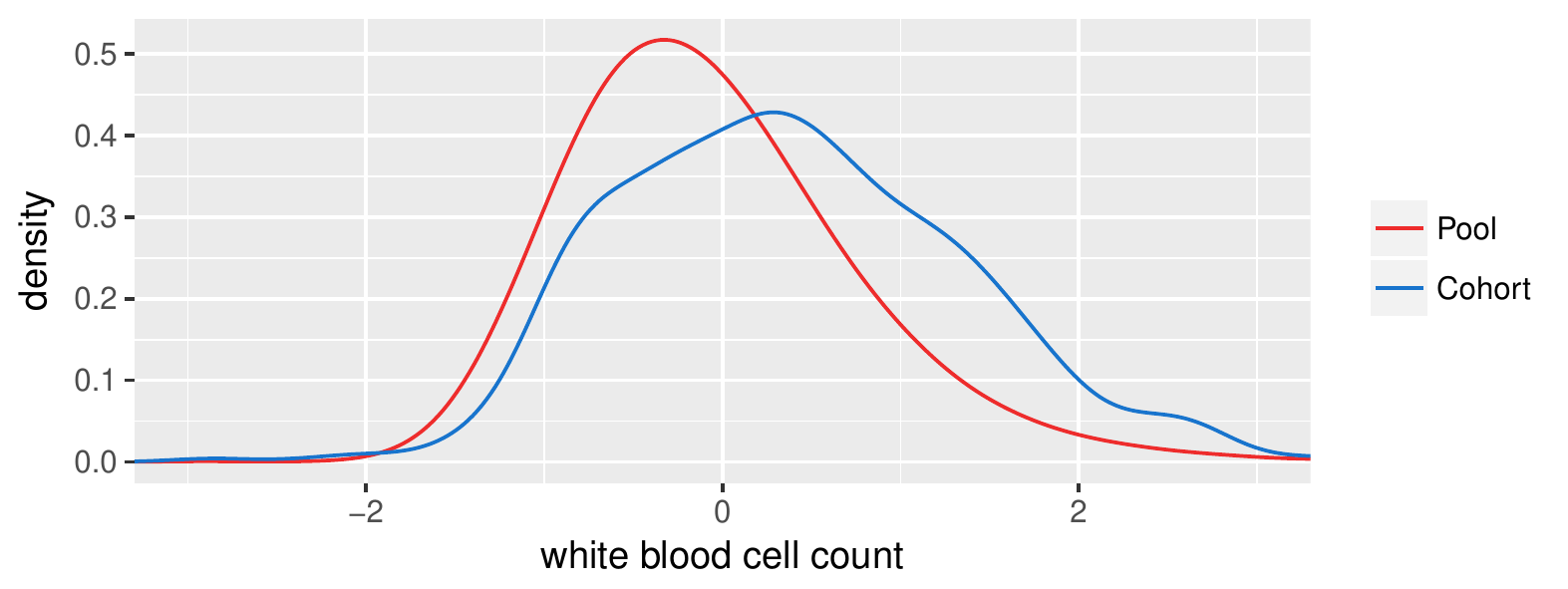}}\\
\end{tabular}
\caption{Empirical densities of continuous covariates in a selectively recruited cohort ($N=1,000$) and the pool ($N=82,089$).}
\end{figure}

%=================================================%
\begin{table}
\centering
\begin{tabular}{|c|c|c|c|c|}
\hline
Covariate & Inferred $\bv$ & Lower CI & Upper CI & p-value \\
\hline
age & 1.274 & 1.252 & 1.295 & 0 \\ 
gender & 0.102 & 0.084 & 0.121 & 1.883265e-27 \\ 
deprived & -0.058 & -0.076 & -0.039 & 8.060814e-10 \\ 
CHD & 0.198 & 0.179 & 0.216 & 1.962481e-97 \\ 
UA & 0.042 & 0.024 & 0.061 & 7.265859e-06 \\ 
NSTEMI & 0.118 & 0.1 & 0.137 & 4.082319e-36 \\ 
STEMI & -0.01 & -0.028 & 0.009 & 0.3110636 \\ 
PCI & -0.067 & -0.085 & -0.048 & 1.220014e-12 \\ 
CABG & -0.134 & -0.152 & -0.115 & 1.291749e-45 \\ 
MI & 0.03 & 0.012 & 0.049 & 0.001271308 \\ 
nitrates & 0.045 & 0.026 & 0.063 & 2.272479e-06 \\ 
smoke & 0.12 & 0.102 & 0.139 & 3.595763e-37 \\ 
hypertension & 0.022 & 0.004 & 0.041 & 0.01787763 \\ 
diabetes & 0.144 & 0.125 & 0.162 & 1.925671e-52 \\ 
chol & 0.001 & -0.023 & 0.024 & 0.9572574 \\ 
hdl & 0.003 & -0.02 & 0.025 & 0.826079 \\ 
heart.failure & 0.248 & 0.229 & 0.266 & 4.505281e-152 \\ 
PAD & 0.169 & 0.151 & 0.187 & 8.125267e-72 \\ 
AF & 0.16 & 0.141 & 0.178 & 2.189484e-64 \\ 
stroke & 0.224 & 0.206 & 0.243 & 4.102543e-125 \\ 
renal & 0.174 & 0.155 & 0.192 & 8.507037e-76 \\ 
COPD & 0.065 & 0.046 & 0.083 & 6.082837e-12 \\ 
cancer & 0.229 & 0.211 & 0.248 & 1.038268e-130 \\ 
liver & 0.48 & 0.462 & 0.499 & 0 \\ 
depression & 0.06 & 0.042 & 0.079 & 1.906276e-10 \\ 
anxiety & -0.065 & -0.084 & -0.047 & 3.798424e-12 \\ 
crea & 0.052 & 0.034 & 0.069 & 3.374309e-09 \\ 
wbc & 0.056 & 0.035 & 0.077 & 1.842707e-07 \\ 
haemo & -0.092 & -0.114 & -0.07 & 9.550895e-17 \\ 
\hline
\end{tabular}
\caption{Inferred parameters from a Cox proportional hazards model applied to the full CALIBER dataset ($N = 82, 089$).}
\label{tab:pool}
\end{table}

%\bibliographystyle{plainnat}
%\bibliography{refs}
\end{document}